\documentclass{elsart}

\usepackage[english]{babel}
\usepackage{dcolumn}
\usepackage{amsmath,amssymb,epsfig,float}

\begin{document}
\begin{frontmatter}
\title{Black hole analogues in braneworld scenario}
\author[apctp]{Xian-Hui Ge}~~~
\author[apctp,kun]{Sung-Won Kim}
\address[apctp]{Asia-Pacific Center for Theoretical Physics,\\
 Pohang 790-784, Korea }
\address[kun]{
Department of Science Education, Ewha Woman's University, Seoul
120-750, Korea}
\thanks[email]{e-mail:gexh@www.apctp.org}
\begin{abstract}
  \hspace*{7.5mm}We construct analogue black hole solutions in the
  braneworld scenario. The quantum fluctuations of condensate
  gravitons
   propagating around a $4+n$-dimensional gravitational
  potential are found yielding a metric similar to higher
  dimensional Schwarzschild black hole line-element.
   Black hole analogue solutions in Randall-Sundrum and Dvali-Gabadadze-Porrati brane world models
  are also constructed. The properties of such black hole analogues
  are discussed.
\end{abstract}
\end{frontmatter}

\section{Introduction}
\hspace*{7.5mm}A lot of interest in recent years has been raised for
field theories where the standard model of high-energy physics is
assumed to live on a 3-brane embedded in a larger space-time, while
only the gravitational fields are in contrast usually considered to
live in the whole spacetime \cite{horava,Arkani,randall,dvali}.
There mainly three kind of pictures in the brane world scenario. The
first picture is proposed by Arkani-Hamed \textit{et al} (ADD model)
\cite{Arkani}, who suggested that the traditional Planck scale,
$M_p$, is only an effective energy scale derived from the
fundamental one in $(4+n)$-dimensional space, $M_{*}$, through the
following relation $M^2_{p}\sim M_{*}^{2+n}R^{n}$, where $R$ is the
size of each extra dimensions.  The fact that we do not see
experimental signs of the extra dimensions despite that the
compactification scale of the extra dimensions  would have to be
much smaller than the weak scale, implies that only gravity can
propagate in the extra-dimensional spacetime and all ordinary
matter: electromagnetic, weak and strong forces, is restricted to
live on a (3+1) dimensional hypersurfaces, a 3-brane.  In the second
picture which we refer as the Randall-Sundrum (RS) model, a new
higher dimensional mechanism for solving the hierarchy problem was
proposed. The weak scale is considered to be generated from the
Planck scale through an exponential hierarchy. A positive tension
3-brane embedded in an 5-dimensional AdS bulk and the cross over
between 4-dimensional and 5-dimensional gravity is set by the AdS
radius\cite{randall}. The third picture is based on the work of
Dvali, Gabadadze, and Porrati (DGP), where the 3-brane is
embedded in 5-dimensional Minkowski space with an infinite size extra dimension.\\
\hspace*{7.5mm} The presence of extra dimensions in brane world
gravity models will inevitably change the properties and physics of
black holes. In ref.\cite{arg}, the authors found that Schwarzschild
black holes in ADD model with horizon  radius smaller than the size
of the new spatial dimensions, $R$, are bigger, colder and
longer-lived than a usual $(3+1)$-dimensional black hole of the same
mass. Physicists expect to observe Tev scale black holes at Large
Hadron Collider (LHC) in the near future \cite{savas}. However,
exact solutions describing a black hole bound to a
3-brane in RS and DGP models is not yet known.\\
  \hspace*{7.5mm}On the other hand, the remarkable work of Unruh, in
1981, developed a way of mapping certain aspects of black holes in
supersonic flows and pointed out that propagation of sound in a
fluid or gas turning supersonic~\cite{unruh}, is similar to the
propagation of a scalar field close to a black hole, and thus
experimental investigation of the Hawking radiation is possible.
From then on, several candidates have been considered for the
experimental test of the analogue of
black holes~\cite{novello,visser}.\\
\hspace*{7.5mm}In the present study, we construct black hole
analogue metrics in the brane world models and investigate their
properties and physics. In ref.\cite{randall}, the authors studied
the linearized gravitational fluctuations in the 5-dimensional space
time while the wave equation is analogous to non-relativistic
quantum mechanics equation. Then they computed the effective
non-relativistic gravitational potential between two particles. In
this paper, we find that the
 non-relativistic
quantum mechanics equation to describe the propagation of
gravitational waves and discuss quantum fluctuations of the
propagation of gravitational waves. In this way, we obtain the
metric of black hole analogues and then study their behaviors in
ADD, RS and DGP models.
\section{General formalisms}
Without loss of generality, we consider a system of Bose-Einstein
condensate particles, which may not  be the standard model particles
and are permitted to propagate in an arbitrary dimensional
space-time ( gravitons, scalar particles etc.). A Bose-Einstein
condensate is the ground state of a second quantized many body
Hamiltonian for $N$ bosons trapped by an external potential
$V(\vec{x})$. When the temperature is low enough, most of all the
particles can be described by the same single-particle quantum state
$\psi (\vec{x},t)$ and the interactions between particles become
sufficiently small. In this sense, the evolution of $\psi
(\vec{x},t)$ is described by the
 Schr$\textrm{{\"{o}}}$dinger equation
 in $(4+n)$-dimensions,
\begin{equation}
 \label{schrod}
[-\frac{\hbar ^2}{2m}\nabla ^2 +V(\rho,\vec{x})]\psi
   (\vec{x},t)=i\hbar\frac{\partial}{\partial t}\psi (\vec{x},t),
   \end{equation}
 where $m$ is the mass of the particles, $\psi (\vec{x},t)$ can be considered as a
   macrostate of condensate scalar particles ( gravitons etc.),
  $V(\rho,\vec{x})$ is a potential in
  $(4+n)$-dimensions and we normalize the total number of particles $\int d^3\textbf{x}|\psi(\vec{x},t)|^2=N$. The mass of
   particles and the total number of particles has the following relation $m=Nm'$, where
  $m'$ is mass per particle.
  In general,
 $V(\rho,\vec{x})$ in quantum mechanics can be electric Coulomb potential
 or some potentials else. We will specify $V(\rho,\vec{x})$ to be the effective
non-relativistic gravitational potentials in brane world models
lately.\\
\hspace*{7.5mm}The Lagrangian
   corresponding to the Schr$\ddot{o}$dinger equation can be
   defined as $\mathcal{L}= {i\hbar}\psi ^{*}\frac{\partial \psi}
   {\partial t}-
   \frac{\hbar^2}{2m}\nabla \psi ^{*}\cdot \nabla
   \psi-V(\vec{x})\psi^{*}\psi$, which can be further rewritten as
   \begin{equation}
   \label{l}
   \mathcal{L}=-\psi^{*}\psi\left\{\frac{\hbar}{i}\frac{\partial}{\partial t}\textrm{ln}\psi+
   \frac{\hbar^2}{2m}\nabla(\textrm{ln}\psi ^{*})\cdot \nabla(\textrm{ln}\psi )+V(\rho,\vec{x})\right\}.\end{equation}
   Comparing Eq.(\ref{l}) with the Jacobi-Hamilton equation, i.e.
   \begin{equation}
   \label{jacobi}
   \frac{\partial S}{\partial t}+\frac{(\nabla S)^2}{2m}+V(\rho,\vec{x})=0,\end{equation}
   where S is the action of the whole system, one obtains
   $S=\frac{\hbar}{i}\rm{ln}\psi$ and $S^{*}=-\frac{\hbar}{i}\rm{ln}\psi^{*}$.
   Assumed $S=S_{r}+i S_{i}$, $\psi$ can be rewritten as
   \begin{equation}
   \label{psi}
   \psi (\vec{x},t)=e^{\frac{i(S_{r}+i S_{i})}
   {\hbar}}=\rho^{1/2} (\vec{x},t)e^{\frac{i
   S_{r}(\vec{x},t)}{\hbar}},\end{equation}
   where $\rho=\psi\psi^{*}$ is  the probability density. The total particle number is then given by $N=\int\rho ~d^3\textbf{x}$.
   Substituting Eq.(\ref{psi}) into the Schr$\ddot{o}$dinger equation, we
   have
   \begin{eqnarray}
   \label{continu}
   &&\frac{\partial \rho}{\partial t}+\nabla \cdot
   \vec{j}(\vec{x},t)=0,\\
   \label{S}
   &&\frac{\partial S_{r}}{\partial t}+\frac{(\nabla S_{r})^2}{2m}+V(\rho,\vec{x})
   -\frac{\hbar^2}{2m}\frac{\nabla ^2 \rho^{1/2}}{\rho^{1/2}}=0,\end{eqnarray}
   where $\vec{j}(\vec{x},t)=\frac{\rho}{m}\nabla S_{r}$ and the last term in Eq.(\ref{S})
    corresponds to the quantum effect without
    classical correspondence.
      We may drop out this
   term by assuming that this term is small compare to other terms and then obtain Euler equation for classical liquid.
   If we set $\vec{v}=\nabla S_{r}/m$, then Eq.(\ref{S}) can be
   rewritten as
   \begin{equation} \frac{\partial \vec{v}}{\partial t}
   +\nabla(\vec{v}^2/2)=-\nabla V(\rho,\vec{x})/m,\end{equation}
   which is the exact equation of irrotational fluid.
   Defining $\Phi=S_{r}/m$, we have
   \begin{equation}
   \label{V}
   \frac{\partial \Phi}{\partial t}+\vec{v}^2/2=
   -V(\rho,\vec{x})/m.
   \end{equation}
    By further defining $\xi=ln \rho$, we have the new forms of Eq.(\ref{continu}) and Eq.(\ref{V})
   \begin{eqnarray}
\label{xi}
   \partial \xi/\partial
   t+\vec{v}\cdot\nabla\xi+\nabla \cdot \vec{v}=0,\\
   \label{par}
   \frac{\partial \Phi}{\partial
   t}+\vec{v}^2/2+ V(\xi,\vec{x})/m=0.\end{eqnarray}
    Linearizing
   Eq.(\ref{xi}) and Eq.(\ref{par}) around the assumed background ($\xi_{0}$,
   $\Phi_{0}$), with $\xi=\xi_{0}+\tilde{\xi}$,
   $\Phi=\Phi_{0}+\tilde{\Phi}$ and $V(\xi,\vec{x})=V(\xi_{0},\vec{x})+V'(\xi_{0},\vec{x})\tilde{\xi}+...$~, we obtain
   \begin{equation}
   \rho_{0}^{-1}\left[\partial \rho_{0}\tilde{\xi}/\partial
   t+\nabla\cdot (\rho_{0}\vec{v}\tilde{\xi}+\rho_{0}\nabla \tilde{\Phi})\right]=0,
   \partial \tilde{\Phi}/\partial t+\vec{v}\cdot\nabla\tilde{\Phi}
   +\tilde{\xi} V'(\xi_{0})/m_{0}=0,\end{equation}
   which result in an equation for $\tilde{\Phi}$,

   \begin{eqnarray}&&\rho_{0}^{-1}\left\{\frac{\partial}{\partial t}
   \left(\frac{m_{0} \rho_{0}}{V'(\xi_{0})}\frac{\partial \tilde{\Phi}}{\partial t}
   +\frac{m_{0}\rho_{0}\vec{v}_{0}}{V'(\xi_{0})) }\cdot \nabla\tilde{\Phi}\right)
\right. \nonumber\\
  &&\left.
  +\nabla\cdot\left(\frac{m_{0}\rho_{0}\vec{v_{0}}}{V'(\xi_{0})}\frac{\partial \tilde{\Phi}}{\partial t}
   -\rho_{0}\nabla \tilde{\Phi} +\vec{v}_{0}\frac{m_{0}\rho_{0}}{V'(\xi_{0})}
   \vec{v}_{0}\cdot\nabla \tilde{\Phi}\right)\right\}=0,\end{eqnarray}
   where $m_{0}=m'\int\rho_{0}d^3\textbf{x}$.
   The above equation is identified with a massless scalar field
   equation describing the sound wave in the curved spacetime
   background,
   \begin{equation}\frac{1}{\sqrt{-g}}\partial_{\mu}(\sqrt{-g}g^{\mu\nu}\partial_{\nu}\tilde{\Phi})=0\end{equation}
with the background metric,
$g_{\mu\nu}=\left(\frac{\rho_{0}}{c}\right)\left(\begin{array}{cc}
  -c^2 +v_{0}^{2} & -v_{0}^{i}  \\
  -v_{0}^{j}& \delta_{ij}
 \end{array}\right)$, which is a $(n+4)\times (n+4)$ matrix.
   The local speed of sound is defined by $c^2\equiv
   |V'(\xi_{0})/m_{0}|$. The
   metric can be written as $ds^2=\frac{\rho_{0}}{c}[(c^2-v_{0}^2)dt^2+2\vec{v}_{0}\cdot d\vec{x}dt-d\vec{x}\cdot
   d\vec{x}]$. Assuming that the background flow is a spherically
   symmetric,~stationary, and convergent flow,we can define a new
   time coordinate by $d\tau=dt+\frac{\vec{v}_{0}\cdot
   d\vec{r}}{c^2-v^2}$. Substituting this back into the line
   element gives
   \begin{equation}
   \label{acousticmetric}
   ds^2=\frac{\rho_{0}}{c}\left[-(c^2-v_{0}^2)d\tau^2+\frac{c^2}{c^2-v_{0}^2}dr^2
   +r^{2}d\Omega^{2}_{n+2}\right],\end{equation}
   This metric denotes the physics of analogue black holes--solid or
   liquid systems that trap scalar particle waves in a way similar to real black holes.
   Here, $c$ plays the role of escaping velocity just as what the speed of light
   means to real black holes. The horizon of the analogue black hole is
   defined by $c^2=v_{0}^2$. Comparing Eq.(\ref{acousticmetric})
   with the 4-dimensional Schwarzschild metric,
\begin{equation}
   \label{schimetric}
   ds'^2=-(c^2-\frac{2GM}{r})dt^2+(1-\frac{2GM}{c^2r})^{-1}dr^2
   +r^{2}d\Omega^{2}_{2}.\end{equation}
If we assume $\frac{c}{\rho_{0}}ds^2=ds'^2$ and
$v_{0}^2=\frac{2GM}{r}$, they identify with each other.
   In the following , we will see that Eq.(\ref{acousticmetric})  describing such a gravitational wave trap are similar to
    those for a black hole's funnel-shaped space time.
    Once such black hole analogues are formed, they should emit Hawking-like radiation from the
    edge of the horizon in the form of phonons. Analogues of black holes thus can be used to detect extra
    dimensions and even study the nature of quantum gravity. In the
    following, we will specify our discussions about the potential
    on gravitational potentials, since gravitons are the only known
    particles that can propagate in extra dimensions \cite{Arkani}.

\section{Black hole analogues in ADD model}
The  gravitational potential in higher dimensional space time can
derived from the equation $\nabla^2 V_{(4+n)}=4\pi G_{(4+n)}\rho$.
We integrate  both sides of equation and use the divergence theorem
on the left-hand and note that the volume integral on the right-hand
side gives the total mass $M$.
 Thus, in ADD model, the gravitational potential for two test particles
 with mass $M$ and $m_{0}$ within a distance  $r\ll R$, is given
 by
 \begin{equation}
V(\xi_{0},\vec{x})= -\frac{4\pi G_{(4+n)}M
m_{0}}{(n+1)r^{n+1}\Omega_{n+2}},
 \end{equation}where $G_{(4+n)}$ denotes the $(4+n)$-dimensional
 Newton constant, and $\Omega_{n+2}=\frac{2\pi^{\frac{n+3}{2}}}{\Gamma(\frac{n+3}{2})}$
 denotes the volume of a unit $(n+2)$ sphere.
If we further assume the velocity $v$ is time independent, then
Eq.(\ref{V}) indicates that $v_{0}^2=\frac{8\pi G_{(4+n)}M
}{(n+1)r^{n+1}\Omega_{n+2}}$. The metric can be written as,
\begin{equation}
\label{add} ds^2=\frac{\rho_{0}}{c}\left[-c^2f(r)d\tau^2
+f^{-1}(r)dr^2+r^{2}d\Omega^{2}_{n+2}\right],
\end{equation} where $f(r)=\left(1-\frac{8\pi G_{(4+n)}M
}{(n+1)r^{n+1}c^2\Omega_{n+2}}\right)$. The above metric is
conformal to Schwarzschild black hole metric in $(4+n)$ dimensions
(only the coefficient is different)\cite{myers}. We should notice
that when $v_{0}$ is chosen, the continuity equation $\nabla \cdot
(\rho_{0}\vec{v}_{0})=0$ then it implies that $\rho_{0}\propto
r^{-\frac{n+3}{2}}$. The presence of the conformal factor does not
influence the properties of black hole analogues much in that the
surface gravity and Hawking temperature are conformal
invariants\cite{jacobson}. If we mainly consider the region near the
event horizon, the conformal factor $\frac{\rho_{0}}{c}$ can simply
be regarded as a constant, and we can ignore the contribution of the
factor \cite{visser}. Hereafter, we focus our discussions on the
near-horizon region and set $\frac{c}{\rho_{0}}ds^2=ds'^2$. The
properties of higher dimensional Schwarzschild black holes in ADD
model have been discussed by Aryres, Dimopoulos and
March-Russell\cite{arg}. We would like to review their main results
here. The event horizon radius is given by
\begin{equation}
\label{hn}
r_{H(4+n)}=\left(\frac{8\pi G_{(4+n)}M
}{(n+1)c^2\Omega_{n+2}}\right)^{\frac{1}{n+1}}.
\end{equation}
Note that in ADD model, $G_{(4+n)}=\frac{1}{M_{*}^{2+n}}$. Thus, we
find that
\begin{equation}
\label{hn2} r_{H(4+n)}\sim
\frac{1}{M_{*}}(\frac{M}{M_{*}})^{\frac{1}{n+1}}
\end{equation}
 When $n=0$, we obtain the four dimensional horizon
radius,
\begin{equation}
r_{H(4)}=(\frac{2G_{(4)}M}{c^2})\sim
\frac{1}{M_{p}}(\frac{M}{M_{p}})\sim \frac{M}{M^{2+n}_{*}R^{n}},
\end{equation}
where the relation  $G^{-1}_{4}\sim M^2_{p}\sim M_{*}^{2+n}R^{n}$
has been used. As a consequence, we have
\begin{equation}
\frac{r_{H(4)}}{r_{H(4+n)}}\sim
\left(\frac{r_{H(4+n)}}{R}\right)^{n}<1,
\end{equation}
Therefore, for small $(4+n)$-dimensional black holes that can
submerge into extra dimensions, the horizon radius is bigger than
that of corresponding $4$-dimensional black holes of the same mass,
i.e.~$r_{H(4)}<r_{H(4+n)}<R$. However, the Hawking temperature
$T_{H(4+n)}$ of a small $(4+n)$-dimensional black hole is found to
be colder than its $4$-dimensional correspondence. The temperature
of black hole analogues can be easily estimated from the standard
formula of acoustic black hole temperature,
\begin{equation}
T_{H(4+n)}=\frac{\hbar}{2\pi
k_{B}}\left|\frac{\partial}{\partial{r}}v_0\right|_{r=r_{H(4+n)}}\sim
\frac{1}{r_{H(4+n)}},
\end{equation}
which is smaller compared with the 4-dimensional temperature
$T_{H(4)}\sim 1/r_{H(4)}$. For such larger and colder higher
dimensional black holes, the lifetime is correspondingly longer than
that of an equal mass 4-dimensional one\cite{arg}. The entropy of
such small black hole analogues is given by
\begin{equation}
\label{entropy}
S_{H(4+n)}=\frac{A_{n+2}}{4G_{(4+n)}}=\frac{r^{n+2}_{H(4+n)}\Omega_{n+2}}{4G_{(4+n)}},
\end{equation} while the 4-dimensional black hole entropy goes as,
\begin{equation}
\label{entro} S_{H(4)}=\frac{r^{2}_{H(4)}\Omega_{2}}{4G_{(4)}}.
\end{equation}
Comparing Eq.(\ref{entropy}) with Eq.(\ref{entro}), we obtain,
\begin{equation}
\label{ent} \frac{S_{H(4+n)}}{S_{H(4)}}\sim
\left(\frac{R}{r_{H(4+n)}}\right)^{n}>1.
\end{equation}
This confirms us that for horizon radius $r_{{H(4)}}<r_{H(4+n)}<R$,
the $(4+n)$-dimensional entropy is larger than the 4-dimensional
one.
\\
\hspace*{7.5mm} In summary, we are able to derive an analogue black
hole metric similar to that of $(4+n)$-dimensional Schwarzschild
black holes by considering the propagations of gravitational waves
around a higher dimensional gravitational potential. The quantum
fluctuations of the propagations of gravitational waves yield a
massless scalar field
 equation in curved space-time. The properties and physics of such
 black hole analogues are just what described in \cite{arg}
 : for horizon radius smaller then the size of extra dimensions, such
 black holes are bigger, colder and longer lifetime than their
 4-dimensional correspondences of the same mass.

\section{ Black hole analogues in RS model}
\hspace*{7.5mm}The gravitational potential in RS models behaves as,
\begin{eqnarray}
\label{rs}
&&V(\vec{x})=-\frac{G_{(4)}Mm}{r}\left(1+\frac{2\ell^2}{3r^2}\right),
~~~~\rm
{for}~~~~ r\gg \ell, \\
&&\rm{and}\\
&&V(\vec{x})=-\frac{G_{(4)}Mm\ell}{r^2},~~~~~~~~~~~~~~\rm{for}~~~~
r\ll \ell
\end{eqnarray} where $\ell$ is the effective size of extra dimension
probed by the 5-dimensional gravitons. In the RSI scenario, there
are two branes (with equal but opposite tensions) separated by a
piece of AdS space between them. The effective scale on the visible
brane is expressed as,
\begin{equation}
M^{2}_{p}=M^{3}_{*}\ell[1-e^{-2L/\ell}].
\end{equation} In the RSII model, there is only one positive tension
brane which can be thought of arising from the negative tension
brane off to infinity, $L\rightarrow \infty$. The energy scales are
then written as
\begin{equation}
M^{2}_{p}=M^{3}_{*}\ell.
\end{equation} We will concentrate mainly on the RSII model and
discuss the properties of RS black hole analogues only for region
$r\gg \ell$, since for $r\ll \ell$ the black hole properties is just
a special case (the case when $n=1$) of what we have discussed in
the last section. Eq.(\ref{rs}) gives the small corrections to
4-dimensional gravity at low energy from extra-dimensional effects.
These effects serve to slightly
strengthen the gravitational field.\\
\hspace*{7.5mm}We  assume the velocity $v$ is time independent, then
Eq.(\ref{V}) indicates that
$v^2_0=\frac{2G_{(4)}M}{r}\left(1+\frac{2\ell^2}{3r^2}\right)$. The
line element of black hole analogues in RS models are then given by,
\begin{equation}
ds^2=\frac{\rho_0}{c}\left[-c^2\left(1-\frac{2G_{(4)}M}{rc^2}(1+\frac{2\ell^2}{3r^2})\right)d\tau^2
+\left(1-\frac{2G_{(4)}M}{rc^2}(1+\frac{2\ell^2}{3r^2})\right)^{-1}dr^2+r^2d\Omega_{3}\right]
\end{equation}
The horizon radius is
$r_{RS}=\frac{2}{3}\frac{G_{(4)}M}{c^2}+\frac{(2G_{(4)}M)^{5/3}}{3c^2a^{1/3}}+\frac{(2G_{(4)}Ma)^{1/3}}{3c^2}$,
where
$a=9c^2\ell^2+4G^{2}_{(4)}M^2+3c^2\ell\sqrt{9c^2\ell^2+8G^{2}_{(4)}M^2}$.
The value of this radius is smaller than the 4-dimensional
Schwarzschild horizon radius of the same mass. The temperature of
black hole analogues in RS models is simply defined by,
\begin{equation}
T_{RS}=\frac{\hbar}{2\pi
k_{B}}\left|\frac{\partial}{\partial{r}}v_0\right|_{r=r_{RS}}=\frac{\hbar}{4\pi
k_{B}}\left(\frac{1}{r_{RS}}+\frac{8G_{(4)}M\ell^2}{3r^{4}_{RS}}\right)
\end{equation}
Note that $r_{RS}<r_{H(4)}$, thus the temperature of RS black hole
analogues for $r\gg l$ is hotter than that of a 4-dimensional
Schwarzschild black hole of the same mass. We can understand
intuitively that the smaller black hole horizon radius is, the
hotter
the temperature becomes.\\
\hspace*{7.5mm} The lifetime of such RS black hole analogues is
correspondingly shorter than that of an equal mass 4-dimensional
Schwarzschild one. The rate at which energy is radiated for RS black
hole analogues is of order,
\begin{equation}
\frac{dE}{d\tau}\sim A_{RS}T^5_{RS},
\end{equation}
where $A_{RS}$ denotes the area of the 5-dimensional RS black hole
analogues. Remember that, $T_{RS}\sim
\left(\frac{1}{r_{RS}}+\mathcal{O}(1/r_{RS})\right)$, we find that,
\begin{equation}
\frac{dE}{d\tau}\sim \frac{\Delta M}{\Delta\tau_{RS}}\sim
\frac{1}{r^2_{RS}}.
\end{equation}
Comparing with energy decaying rate of a 4-dimensional Schwarzschild
black hole with temperature $T_{H}\sim \frac{1}{r_{H(4)}}$,
\begin{equation}
\frac{\Delta M}{\Delta\tau_{H}}\sim \frac{1}{r^2_{H(4)}},
\end{equation} we find
\begin{equation}
\frac{\Delta\tau_{RS}}{\Delta\tau_{H}}\sim
\frac{r^2_{RS}}{r^2_{H(4)}}<1.
\end{equation}
The entropy of an RS black hole induced on the 3-brane is given by,
\begin{equation}
S_{RS}=\frac{A_{RS}}{4G_{(4)}}=\frac{r^2_{RS}\Omega_{2}}{4G_{(4)}}.
\end{equation}
Once compare with the entropy of an equal mass 4-dimensional
Schwarzschild black hole (see Eq.(\ref{entro})), we find that
\begin{equation}
\frac{S_{RS}}{S_{H(4)}}\sim \frac{r^2_{RS}}{r^2_{H(4)}}<1.
\end{equation}
Therefore, the properties of an RS black hole analogue in the region
$r\gg \ell$, compared with a 4-dimensional Schwarzschild black hole,
is smaller, hotter and shorter-lived, while in the region $r\ll
\ell$, the physics of such black hole analogues becomes
undistinguishable from that of an ADD black hole: bigger, colder and
long-lived than an equal mass 4-dimensional Schwarzschild black
hole.

\section{Black hole analogues in DGP model}
\hspace*{7.5mm}Gravitational potential  in DGP model is the exact
4-dimensional potential at short distances whereas at large
distances the potential is that of a 5-dimensional theory. We mainly
interested in the large distance potential, when $r\gg r_{0}$, we
have,
\begin{equation}
V(\vec{x})=-\frac{G_{(4)}Mmr_0}{r^2},
\end{equation}where $r_0=\frac{M_p^2}{2M^3_{*}}$.
Therefore, the line element for large distance $r\gg r_{0}$ read as,
\begin{equation}
\label{dgp}
ds^2=\frac{\rho_0}{c}\left[-c^2(1-\frac{2G_{(4)}Mr_0}{r^2c^2})d\tau^2+(1-\frac{2G_{(4)}Mr_0}{r^2c^2})^{-1}dr^2+r^2d\Omega^2_3\right]
\end{equation}This solution in fact is a special case of the
black hole metric given in Eq.(\ref{add}) when $n=1$. However, the
physics of black hole analogues  in DGP model are different from
that in ADD model. In ADD model, if the black hole horizon is much
larger then the size of the extra dimensions, $r_{H}\gg R$, the
nature of the black hole is actually a 4-dimensional one. It is only
when $r_{H}\ll R$, the black hole becomes virtually a higher
dimensional object that is submerged into the extra-dimensional
space-time. The DGP model, however, presents us an inverse picture:
when the horizon of a black hole is larger than the size of the
extra dimension, $r_{H}\gg r_0$, the produced black hole is in fact
a five-dimensional one; when $r_{H}\ll r_0$, we recover to the usual
4-dimensional Einstein gravity and the black hole becomes a
4-dimensional object.\\
\hspace*{7.5mm} The horizon radius can be obtained from
Eq.(\ref{dgp}),
\begin{equation}
r_{DGP(5)}=(\frac{2G_{(4)}Mr_0}{c^2})^{1/2}
\end{equation}
 Following the analysis in section 1, one can find that he horizon radius in the region $r_{H}\gg r_0$ of
DGP black hole analogues is larger than that of a 4-dimensional
Schwarzschild black hole with the same mass, namely,
\begin{equation}
r_{DGP(5)} > r_{H(4)}
\end{equation}
The temperature  and life-time of such DGP black hole analogues are
respectively colder and longer-lived than its four-dimensional
correspondences. The entropy is
then larger than that of a 4-dimensional Schwarzschild black hole.\\
\hspace*{7.5mm}In general speaking, the scale of $r_0$ in DGP model
is always believed to be of order of Hubble radius. So we do not
expect to find such DGP black hole analogues in experiments. The
properties of DGP black hole analogues discussed above are only of
theoretical significance.
\section{Conclusions}
 \hspace*{7.5mm} In conclusion, we have constructed
black hole analogue solutions in the braneworld models by
considering the propagation of gravitational waves in
$(4+n)$-dimensional space time. The black hole analogue metric in
ADD model is similar to the higher dimensional Schwarzschild black
hole line-element. Black hole analogues in RS model are also
discussed and it is found that for horizon radius larger than the
size of extra dimension $r_{H}\gg \ell$, the analogues are smaller,
hotter and shorter-lived than that of an equal mass 4-dimensional
Schwarzschild black hole, while in the region $r\ll \ell$, the
physics of such black hole analogues becomes undistinguishable from
that of an ADD black hole. The properties of DGP black hole
analogues are also briefly discussed. When the horizon radius is
larger than the size of the extra dimension $r\gg r_0$, the DGP
black hole analogues are found to be larger, colder and longer-lived
than its
equal mass 4-dimensional correspondence.\\
  \hspace*{7.5mm}The above discussions show that
gravitational wave black holes have similar properties to that of
higher dimensional Schwarzschild black holes. However, experimental
realizing of such black hole analogues in laboratory could be no
less difficult than finding mini black holes at LHC since
gravitational waves are still elusive to the experiments. But this
does not mean that one can preclude such possibilities of detecting
extra dimensions by using gravitational wave black hole analogues.
So far, in the discussions above, we have assumed the interactions
between condensate particles vanished and the local speed sound to
be a constant. In fact, $c$ should be
position-dependent\cite{visser}. It is convenient to discuss the
properties of black hole analogues and to compared with that of real
black holes by assuming $c$ to be a constant.
We need further investigations on these points.\\
\hspace*{7.5mm}The correlations between the black hole
  mass and its temperature, deduced from the energy spectrum of the
  decay products, can test Hawking's evaporation law. The
amplitude probability in the bulk and on the brane allows one to
find its dependence on the number of extra dimensions $n$ and the
angular momentum number $l$. Kanti and March-Russell have calculated
the greybody factor of $(4+n)$-dimensional Schwarzschild for scalar
particles propagating in the bulk and localized on the brane
\cite{kmr}(for related work see also \cite{related}). We would like
to calculate the grey factors of the radiation spectrum for higher
dimensional black hole analogues in our forthcoming paper.

 {\textbf{Acknowledgments}}\\
 S. W.
Kim is supported in part by KRF.

\end{document}